\documentclass[molecules,article,submit,moreauthors,pdftex]{Definitions/mdpi}

\firstpage{1}
\pubvolume{1}
\issuenum{1}
\articlenumber{0}
\pubyear{2021}
\copyrightyear{2020}
\externaleditor{Academic Editor:} 
\datereceived{}
\dateaccepted{}
\datepublished{}
\hreflink{https://doi.org/} 

\Title{Best-practice aspects of quantum-computer calculations: A case study of hydrogen molecule}

\TitleCitation{Best-practice aspects of quantum-computer calculations: A case study of hydrogen molecule}

\Author{Ivana Mih{\' a}likov{\' a}$^{1,2,3}$\orcidA{}, Martin Fri{\' a}k$^{1,5}$\orcidD{}, Matej Pivoluska$^{2,4}$\orcidC{}, Martin Plesch$^{2,4}$*\orcidB{}, Martin Saip$^{5}$\orcidE{} and Mojm{\' i}r {\v S}ob$^{5,1}$\orcidG{}}

\AuthorNames{Ivana Mih{\' a}likov{\' a}, Martin Fri{\' a}k, Matej Pivoluska, Martin Plesch, Martin Saip and Mojm{\' i}r {\v S}ob}

\AuthorCitation{Mih{\' a}likov{\' a}, I.; Fri{\' a}k, M.; Pivoluska, M.; Plesch, M.; Saip, M.; {\v S}ob, M.}

\address{%
$^{1}$ \quad Institute of Physics of Materials, v.v.i., Czech Academy of Sciences, {\v Z}i{\v z}kova 22, CZ-616 62 Brno, \\ Czech Republic; mihalikova@ipm.cz, friak@ipm.cz\\
$^{2}$ \quad Institute of Computer Science, Masaryk University, {\v S}umavsk{\' a} 416, CZ-602 00 Brno, Czech Republic; \\
$^{3}$ \quad Department of Condensed Matter Physics, Faculty of Science, Masaryk University, Kotl{\' a}{\v r}sk{\' a} 2, \\ CZ-611 37 Brno, Czech Republic; \\
$^{4}$ \quad Institute of Physics, Slovak Academy of Sciences, D{\' u}bravsk{\' a} cesta 9, SK-841 04 Bratislava, Slovak Republic; \\
$^{5}$ \quad Department of Chemistry, Faculty of Science, Masaryk University, Kotl{\' a}{\v r}sk{\' a}   2, CZ-611 37 Brno, Czech Republic;
}

\corres{Correspondence: martin.plesch@savba.sk}

\abstract{Quantum computers are reaching one crucial milestone after another. Motivated by their progress in quantum chemistry, we have performed an extensive series of simulations of quantum-computer runs that were aimed at inspecting best-practice aspects of these calculations. In order to compare the performance of different set-ups, the ground-state energy of hydrogen molecule has been chosen as a benchmark for which the exact solution exists in literature. Applying variational quantum eigensolver (VQE) to a qubit Hamiltonian obtained by the Bravyi-Kitaev transformation we have analyzed the impact of various computational technicalities. These include (i) the choice of optimization methods, (ii) the architecture of quantum circuits, as well as (iii) different types of noise when simulating real quantum processors. On these we eventually performed a series of experimental runs as a complement to our simulations. The SPSA and COBYLA optimization methods have clearly outperformed the Nelder-Mead and Powell methods. The results obtained when using the $R_{\mathrm{y}}$ variational form were better than those obtained when the $R_{\mathrm{y}}R_{\mathrm{z}}$ form was used. The choice of an optimum {entangling layer} was sensitively interlinked with the choice of the optimization method. The circular {entangling layer} has been found to worsen the performance of the COBYLA method while the full {entangling layer} improved it. All four optimization methods sometimes lead to an energy that corresponds to an excited state rather than the ground state. We also show that a similarity analysis of measured probabilities can provide a useful insight.}


\begin{document}
\end{paracol}

\section{Introduction}

Quantum computing has recently emerged as a very promising alternative to conventional computational means. Conventional supercomputers, albeit versatile and remarkably reliable, seem to be outpaced by ever increasing demand for computational power when developing new drugs~\cite{drug-discovery}, modeling nanoparticles~\cite{Polsterova-2020} or assessing problems in materials science~\cite{materials-science} and nuclear physics~\cite{nuclear-physics1,nuclear-physics2}. In contrast to the well established conventional technologies, quantum computers are expected to provide exponentially growing computational power thanks to the their use of quantum effects~\cite{Feynman,10.5555/1972505} and first indications of so-called quantum advantage/supremacy have already been
demonstrated~\cite{PhysRevLett.127.180501,Arute2019,PhysRevLett.127.180502}. Unfortunately, the current capabilities of quantum computers are still rather limited by numerous methodological issues, lack of suitable software tools, challenges when physically realizing quantum circuits, noise that is reducing their reliability, as well as a very low number of quantum platforms available for users.


Despite the above mentioned challenges, the onset of quantum computers is undeniable and quantum chemistry is one of the most active areas of their applications~\cite{Friesner-2005,Trygve-2008,Cremer-2011,Lyakh-2012,Narbe-2017}. In particular, quantum-mechanical calculations of properties of small molecular systems represent one of the most successful utilization of quantum calculations~\cite{Preskill2018quantumcomputingin,OMalley-PRX-2016,SQ,Cao2019}.
Importantly, quantum computers are no different from their classical counterparts regarding numerous technicalities that are critical for their successful performance~\cite{kandala}. The architecture of quantum circuits, optimization methods used to reach the energy minimum, and numerous computational parameters  critically affect the calculations.

Our study aims at identifying the impact of different computational set-ups and we will use the ground-state energy of H$_2$ molecule as a benchmark for which the exact solution exists. {Such an initial testing of set-ups is} important whenever starting a new set of calculations, e.g., for a new molecular system or a different molecular property.
Our results clearly show that an extensive testing can not only reduce the use of valuable computational resources but can be also critical for obtaining the correct minimum-energy state at all.

\section{Methods}

\begin{figure}[H]
	\centering
\includegraphics[width=0.95\linewidth]{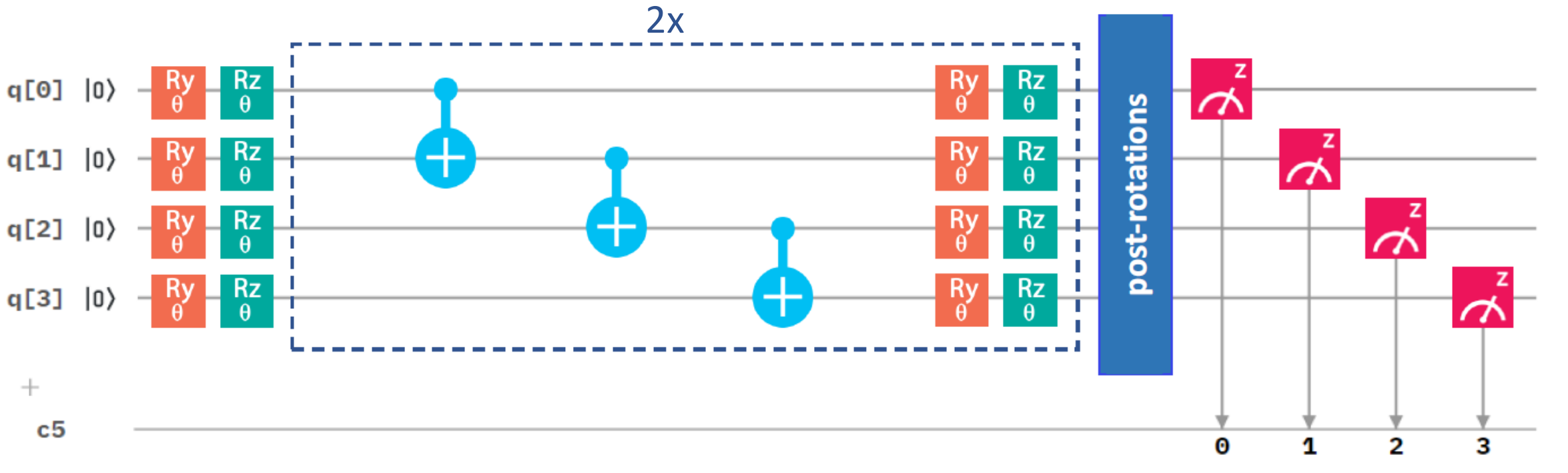}
\caption{A schematics of used quantum circuit with the $R_{\mathrm{y}}R_{\mathrm{z}}$ variational form and the linear {entangling layer} of qubits. It contains gates rotating the qubits (orange and green symbols), control NOT gates responsible for {entangling layers} (light blue symbols) and measurement (red symbol). The central part marked by blue dashed lines is  twice sequentially  in our circuit.}
	\label{fig-form-RyRz}
\end{figure}

The electronic wave functions of the H$_2$ molecule were searched for in  Slater-type orbital basis set with each orbital expanded into 3 Gaussian functions (STO-3G). The distance between the atoms was set to  0.725~\AA \, and the Coulomb repulsion between nuclei is not taken into account. In order to use a quantum computer it is necessary to transform the electronic Hamiltonian of the studied system from its first-quantization form into the second-quantization one~\cite{SQ}.
It is further mapped onto qubit operators represented in the Pauli operator basis~\cite{map} by a suitable transformation, e.g., the Bravyi-Kitaev~\cite{bravyi} {(BK)} or Jordan-Wigner~\cite{Jordan1928} {(JW)} one. We have used the former transformation employing the {\em qiskit} package~\cite{VQETut}. Our code is available in Ref.~\cite{imihalik}.
 A final $4$-qubit Hamiltonian is then
\begin{align}
	\hat{H}_{\mathrm{H_2}}^{\mathrm{4-qubit}}  &= c_{\mathrm 0} \; \boldsymbol{1}
	+ c_{\mathrm 1} \; Z_{\mathrm 0}
	+ c_{\mathrm 2} \;  Z_{\mathrm 1}Z_{\mathrm 0}
	+ c_{\mathrm 1}  \;  Z_{\mathrm 2}
	+ c_{\mathrm 2}  \;  Z_{\mathrm 3}Z_{\mathrm 2}Z_{\mathrm 1}
	+ c_{\mathrm 3}  \;  Z_{\mathrm 1}
	+ c_{\mathrm 4}  \;  Z_{\mathrm 2}Z_{\mathrm 0} \nonumber\\
		&+ c_{\mathrm 5}  \;  X_{\mathrm 2}Z_{\mathrm 1}X_{\mathrm 0}
+ c_{\mathrm 6}  \;  Z_{\mathrm 3}X_{\mathrm 2}X_{\mathrm 0}
	+ c_{\mathrm 6}  \;  X_{\mathrm 2}X_{\mathrm 0}
	+ c_{\mathrm 5}  \;  Z_{\mathrm 3}X_{\mathrm 2}Z_{\mathrm 1}X_{\mathrm 0} \label{H2_4_qubits}\\
	&
	+ c_{\mathrm 7}  \;  Z_{\mathrm 3}Z_{\mathrm 2}Z_{\mathrm 1}Z_{\mathrm 0}
	+ c_{\mathrm 7}  \;  Z_{\mathrm 2}Z_{\mathrm 1}Z_{\mathrm 0}
	+ c_{\mathrm 8}  \;  Z_{\mathrm 3}Z_{\mathrm 2}Z_{\mathrm 0}
	+ c_{\mathrm 3}  \;  Z_{\mathrm 3}Z_{\mathrm 1}. \nonumber
\end{align}
\noindent
The $Z$ and $X$ terms represent Pauli operators and the coefficients $c_i$ are integrals evaluated by the {\em qiskit.chemistry} package \cite{qiskit.chemistry}:
$c_{\rm 0}$ = -0.80718, $c_{\rm 1}$ = 0.17374, $c_{\rm 2}$ =-0.23047, $c_{\rm 3}$ = 0.12149, $c_{\rm 4}$ = 0.16940, $c_{\rm 5}$ = -0.04509, $c_{\rm 6}$ = 0.04509, $ c_{\rm 7}$ = 0.16658, $c_{\rm 8}$ = 0.17511. To solve this Hamiltonian we used one of the most popular quantum-computing approaches, the variational quantum eigensolver (VQE)~\cite{vqe,VQETut,tilly2021variational} involving a quantum circuit such as depicted in Fig.~\ref{fig-form-RyRz}.
{Referring to the VQE description, such as in e.g. the recent review~\cite{tilly2021variational}, the VQE starts with initializing the qubit register (the left-most part of Fig.~\ref{fig-form-RyRz}). }

\newpage

{A quantum circuit is then applied to the qubit register in order to model the physics and entanglement of the electronic wavefunctions.
 A quantum circuit consists of a series of quantum operations on the qubits~\cite{10.5555/1972505}. The structure of the circuit, i.e. a set of ordered quantum gates, is called  an ``ansatz'', and its behavior is defined by a set of parameters (the middle part of Fig.~\ref{fig-form-RyRz}). Once qubits are initialized,  their state is designed to model a trial wavefunction.
The Hamiltonian of the studied system can be measured with respect to this wavefunction to estimate the energy (the right-most part of Fig.~\ref{fig-form-RyRz}). As the measurement of the expectation value is performed only in the computational ($Z$-)basis, the terms in our qubit Hamiltonian containing the Pauli $X$ operators must be measured in a non-diagonal basis. Therefore, basis-switching single-qubit gates (post-rotations) are utilized, see Fig.~\ref{fig-form-RyRz}.
The VQE then variationally optimizes the parameters of the ansatz in order to minimize the trial energy~\cite{Rayleigh,Ritz1909,Rayleigh-Ritz}.
It is a hybrid approach when a conventional computer is used to determine a new set of gates-defining parameters based on the measurements of the quantum circuit employing an classical optimization method.}

Not all of the Pauli terms in our Hamiltonian need to be determined individually as the Pauli operators which require the same post-rotations in the tensor product basis sets can be grouped. Only two circuits (that we will call circuit 0 and 1 below) are needed in our case. Numerous runs (so-called shots) and measurements of probabilities of basis states are needed  to get reliable expectation values for these two circuits (we have used 4096 shots unless specified differently)~\cite{nase-Nanomaterials}.

We have
mostly used simulations of quantum-computer runs, both ideal without noise and noisy ones, but our simulations were complemented also by experiments using real quantum devices. To access them, we used the publicly available cloud-based quantum computing platform IBM Quantum~\cite{IBM}. In order to save these precious resources, we have used only 2-qubit Hamiltonian then. It was possible because the Hamiltonian in Eq.~\ref{H2_4_qubits}
commutes with $Z_1$ and with $Z_3$. Consequently, the Hamiltonian is block-diagonal with $4$ blocks when each of them corresponds to a particular computational basis setting of the qubits 1 and 3. This opens the way to finding a Hamiltonian with the same ground-state energy expressed in a $2$-qubit space~\cite{qiskit.chemistry}:
\begin{equation}
	\hat{H}_{\mathrm{H_2}}^{\mathrm{2-qubit}}
	=  c_{\mathrm 0} \; \boldsymbol{1} + c_{\mathrm 1} \; Z_{\mathrm 0}
	+ c_{\mathrm 1} \; Z_{\mathrm 1}
	+ c_{\mathrm 2}  \; Z_{\mathrm 1}Z_{\mathrm 0}
	+ c_{\mathrm 3}  \; X_{\mathrm 1}X_{\mathrm 0},
    \label{H2_2_qubits}
\end{equation}
\noindent
with the coefficients equal to $c_{\rm 0}$ = -1.05016, $c_{\rm 1}$ = 0.40421, $c_{\rm 2}$ = 0.01135, $c_{\rm 3}$ = 0.18038.

\section{Results}

\begin{figure}[H]
	\centering
\includegraphics[width=0.99\linewidth]{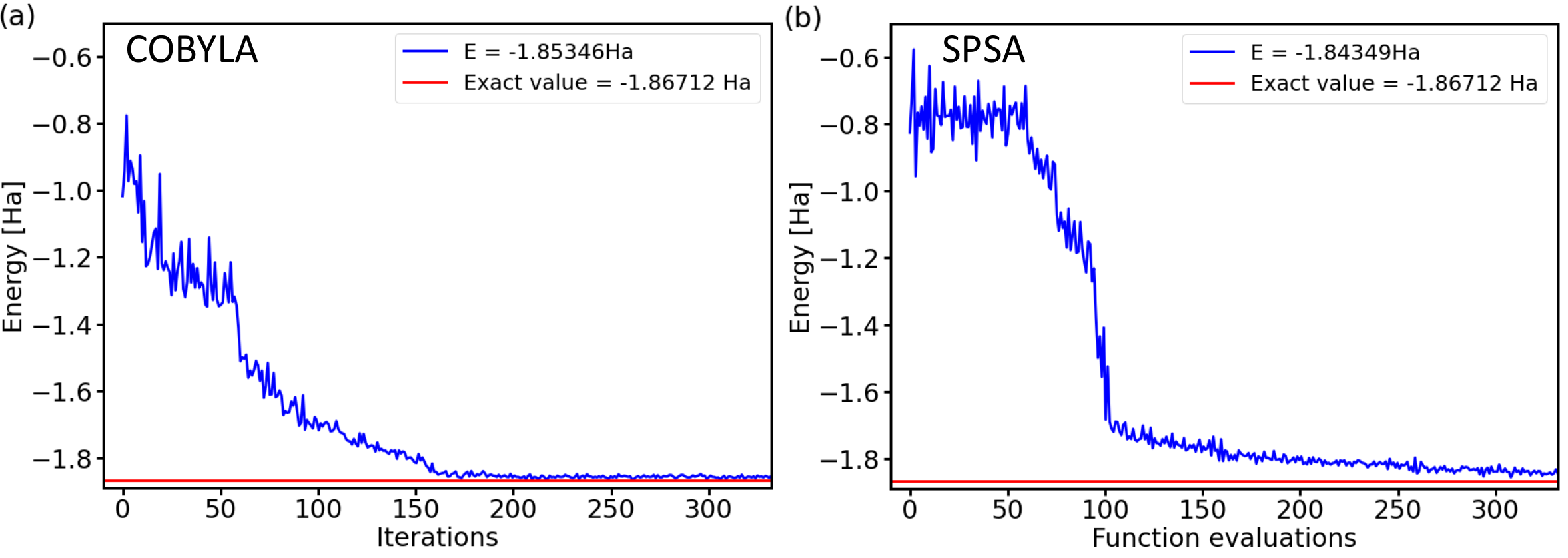}
\caption{Dependencies of the ground-state energies of the H$_2$ molecule as a function of the number of iterations during the optimization process by the classical optimization methods COBYLA (\textbf{a}) and SPSA (\textbf{b})
in the case of a 4-qubit Hamiltonian obtained by the  Bravyi-Kitaev transformation with the circuit architecture characterized by the $R_{\mathrm{y}}R_{\mathrm{z}}$ variational form and the linear {entangling layer} of qubits.}
	\label{fig-41}
\end{figure}

Two examples of optimization process, when the ground-state energy of the H$_2$ molecule is searched for, are presented in Fig.~\ref{fig-41}.

\newpage
The two subfigures in Fig.~\ref{fig-41} present energies as a function of the number of iterations (functional evaluations) for either the
constrained optimization by linear approximation (COBYLA) method~\cite{cobyla} or
the simultaneous perturbation stochastic approximation (SPSA)~\cite{SPSA,Spall1,Spall} optimization method.
The results were obtained for a 4-qubit BK-transformed Hamiltonian with the circuit architecture characterized by the $R_{\mathrm{y}}R_{\mathrm{z}}$ variational form and the linear {entangling layer} of qubits. The comparison of both converging trends is interesting. On the one hand both optimization methods converge to very similar final energies within rather similar number of iterations and, importantly, both converged energies are very close to the exact value. On the other hand Fig.~\ref{fig-41} also clearly shows that the iterative process itself is very different and sensitive to the used optimization method.

The COBYLA optimization is a gradient-free method that provides energies decreasing by smaller amounts but rather monotonously (we leave aside a certain level of numerical noise). In contrast to that, the SPSA method requires a series of initial iterations for the evaluation of pseudo-gradients but once these are determined the energy is decreased more abruptly by a larger amount in a smaller number of iterations.


\subsection{Performance of various optimization methods}

\begin{figure}[H]
	\centering
\includegraphics[width=0.99\linewidth]{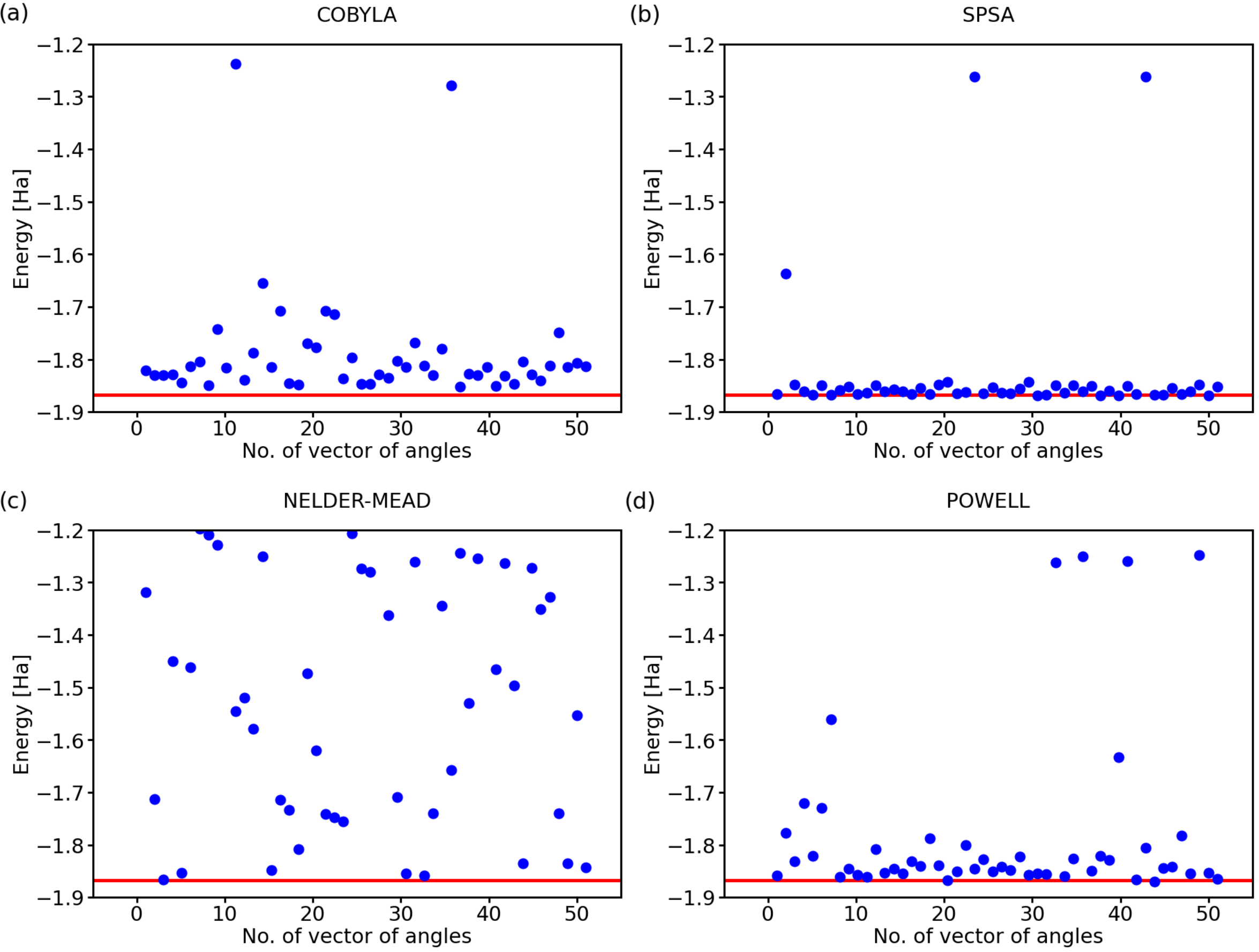}
\caption{Computed ground-state energies of the H$_2$ molecule when using 50 different initial sets of random angles and four classical optimization methods: COBYLA (\textbf{a}), SPSA (\textbf{b}), Nelder-Mead (\textbf{c}) and Powell (\textbf{d}). The employed 4-qubit Hamiltonian was obtained by the Bravyi-Kitaev transformation, the circuit exhibited the $R_{\mathrm{y}}R_{\mathrm{z}}$ variational form, linear {entangling layer} and the evaluations were based on 4096 shots. The red line represents the exact value. }
	\label{fig-42}
\end{figure}

While the above discussed examples are illustrative, it is hard to draw conclusions from a single simulation run for each optimization method. Therefore, we below exhibit results for a series of 50 calculations with randomly selected initial set of angles defining the state of qubits. Further, we analyze also two additional optimization methods, a classical Nelder-Mead~\cite{Nelder-Mead} and that of Powell~\cite{Powell2006}, next to the COBYLA and SPSA. The predicted ground-state energies of hydrogen molecule for 50 simulations using these four optimization methods are shown in Fig.~\ref{fig-42}.

\newpage

 Clearly, there are rather striking differences in the performance
of the analyzed methods. As far as the COBYLA method is concerned, see Fig.~\ref{fig-42}(a), most of the simulations resulted in energies that are quite far away from the exact one covering a range as wide as 0.3 Ha between the exact value (-1.867 Ha) and -1.65 Ha. Two calculations provided energies that are clearly very far away from the ground state of H$_2$ and these energies are, in fact, quite close to  those of excited states of the hydrogen molecule. In particular, the eigenvalues of the 4-qubit Hamiltonian of H$_2$ molecule, as determined using classical techniques, are as follows

$$
\label{eigenvalues}
-1.867, -1.262, -1.262, -1.242, -1.242, -1.242, -1.160, -1.160,
$$
$$
-0.881, -0.465, -0.465, -0.341, -0.341, -0.211, 0.000, 0.227 \ \ \mbox{Ha}.
$$

The values between -1.3 Ha and -1.2 Ha in Fig.~\ref{fig-42}(a) can correspond to excited states with the energy equal to -1.262 Ha or -1.242 Ha.
While these energies of excited states are in principle interesting and physically plausible (as a solution for some valid states of the H$_2$ molecule), we should keep in mind that they represent incorrect predictions as far as our search for the ground-state energy is concerned.

The SPSA method, see Fig.~\ref{fig-42}(b), offers most of the energies very close to the exact value with only two computational runs ending up in the energy region corresponding to excited states and only one simulation providing a value that is clearly erroneous. The Nelder-Mead method, see Fig.~\ref{fig-42}(c), clearly fails to converge to the correct value in a vast majority of cases with only 5-8 simulations, i.e. 10-16 \%, providing the correct energy. While about 20\% of results between -1.2 Ha and -1.3 Ha can be possibly assigned to excited states, all other values are clearly erroneous.

Lastly, the Powell method, see Fig.~\ref{fig-42}(d), represents an intermediate case as far as the accuracy is concerned. While the spread of computed energies of the ground state is smaller than that
obtained for the COBYLA method, the energies are predicted twice more often in the region corresponding to excited states (when compared with the COBYLA results).

\subsection{Impact of the variational form}

\begin{figure}[H]
	\centering
\includegraphics[width=0.99\linewidth]{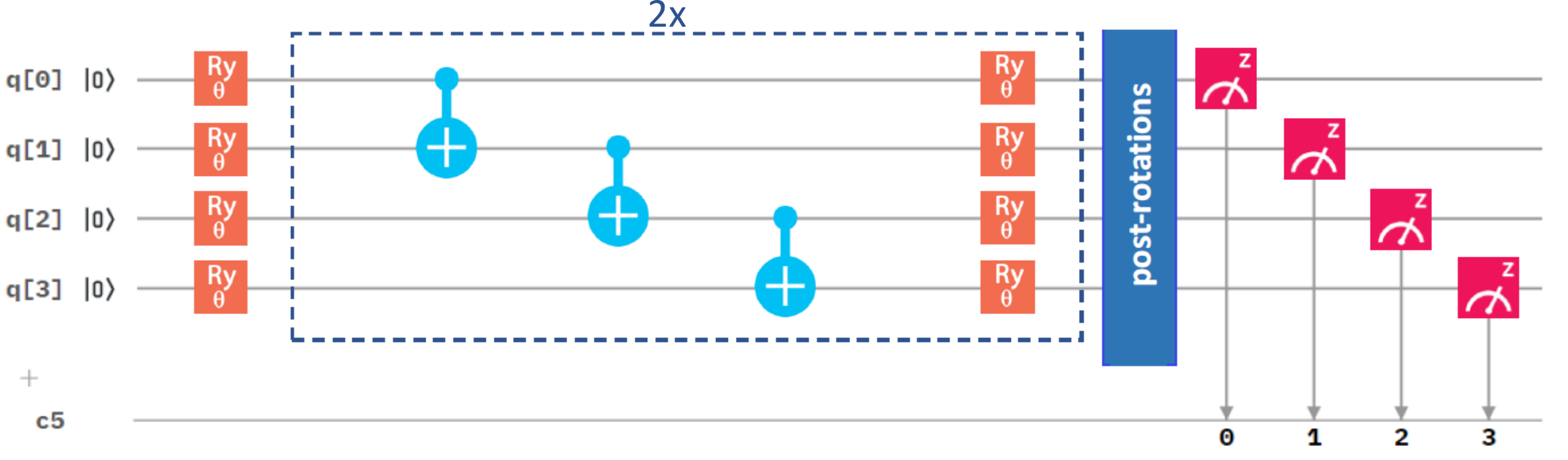}
\caption{A schematic visualization of a circuit containing the $R_{\mathrm{y}}$ variational form and the linear {entangling layer} of qubits. {The central part marked by blue dashed lines is sequentially repeated twice in our circuit.}}
	\label{fig-form-Ry}
\end{figure}

After analyzing the performance of different optimization methods, we next focus on the circuit architecture and we start with the variational form. We compare our initial one in Fig.~\ref{fig-form-RyRz} with an alternative one in Fig.~\ref{fig-form-Ry}.

\begin{figure}[H]
	\centering
\includegraphics[width=0.99\linewidth]{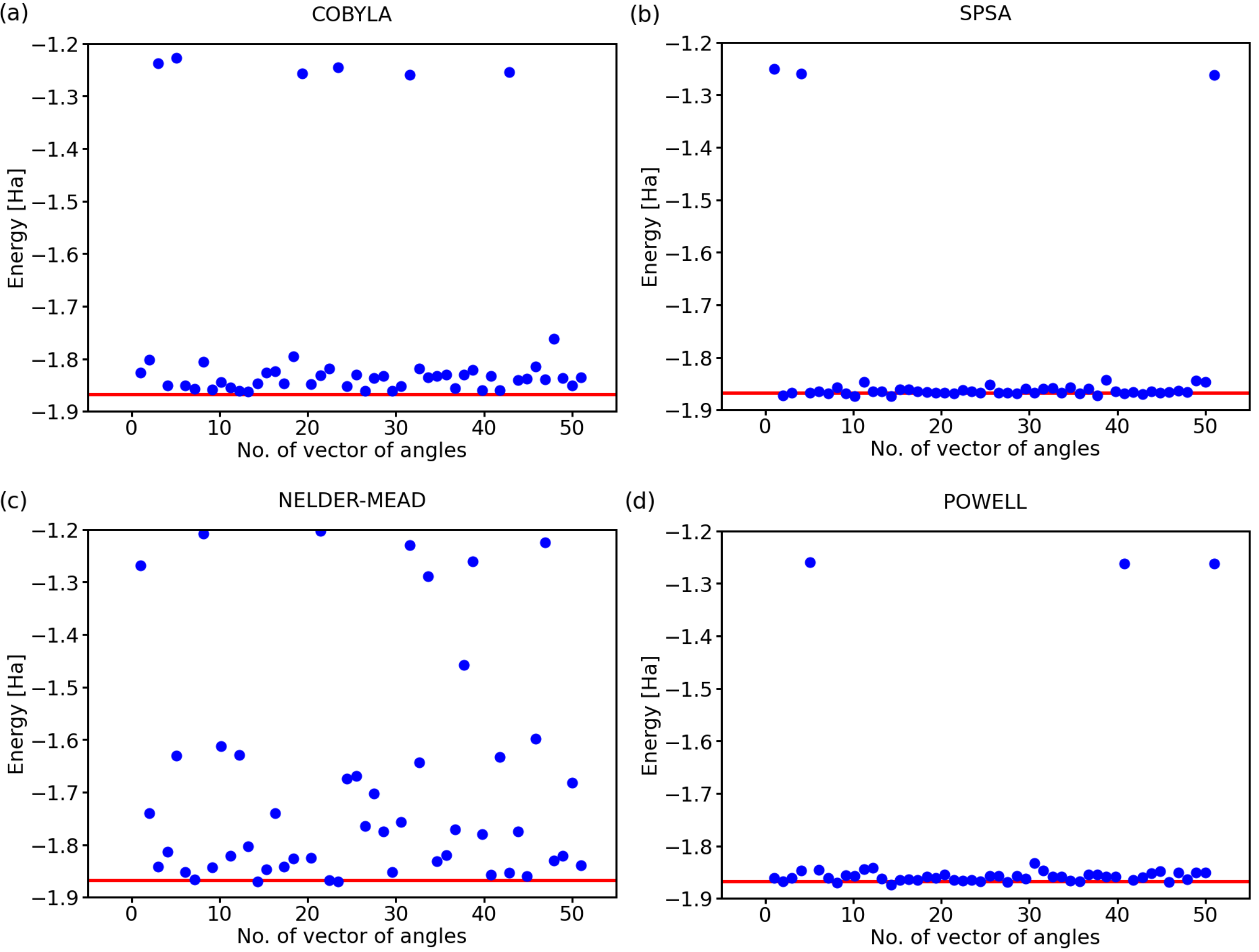}
\caption{
Similar sets of results as in Fig.~\ref{fig-42}, i.e. obtained using classical optimization methods, COBYLA (\textbf{a}), SPSA (\textbf{b}), Nelder-Mead (\textbf{c}) and Powell (\textbf{d}), but the  variational form is the $R_{\mathrm{y}}$ (other circuit characteristics are same as in the case of
 Fig.~\ref{fig-42}).
}
	\label{fig-43}
\end{figure}

They differ in the number of gates. While the results in Fig.~\ref{fig-42} were obtained for the $R_{\mathrm{y}}R_{\mathrm{z}}$ variational form, see Fig.~\ref{fig-form-RyRz},  Fig.~\ref{fig-43} summarizes a corresponding set of results in the case of the $R_{\mathrm{y}}$ variational form (that is above shown in Fig.~\ref{fig-form-Ry}).
The comparison of results in Fig.~\ref{fig-42} and Fig.~\ref{fig-43} clearly shows that the use of the $R_{\mathrm{y}}$ variational form leads to more accurate results even in the case of the Nelder-Mead method. We will, therefore, use the $R_{\mathrm{y}}$ variational form in our simulations below unless specified otherwise.

\newpage


\subsection{Influence of details of the {entangling layers}}

\begin{figure}[H]
	\centering
\includegraphics[width=0.99\linewidth]{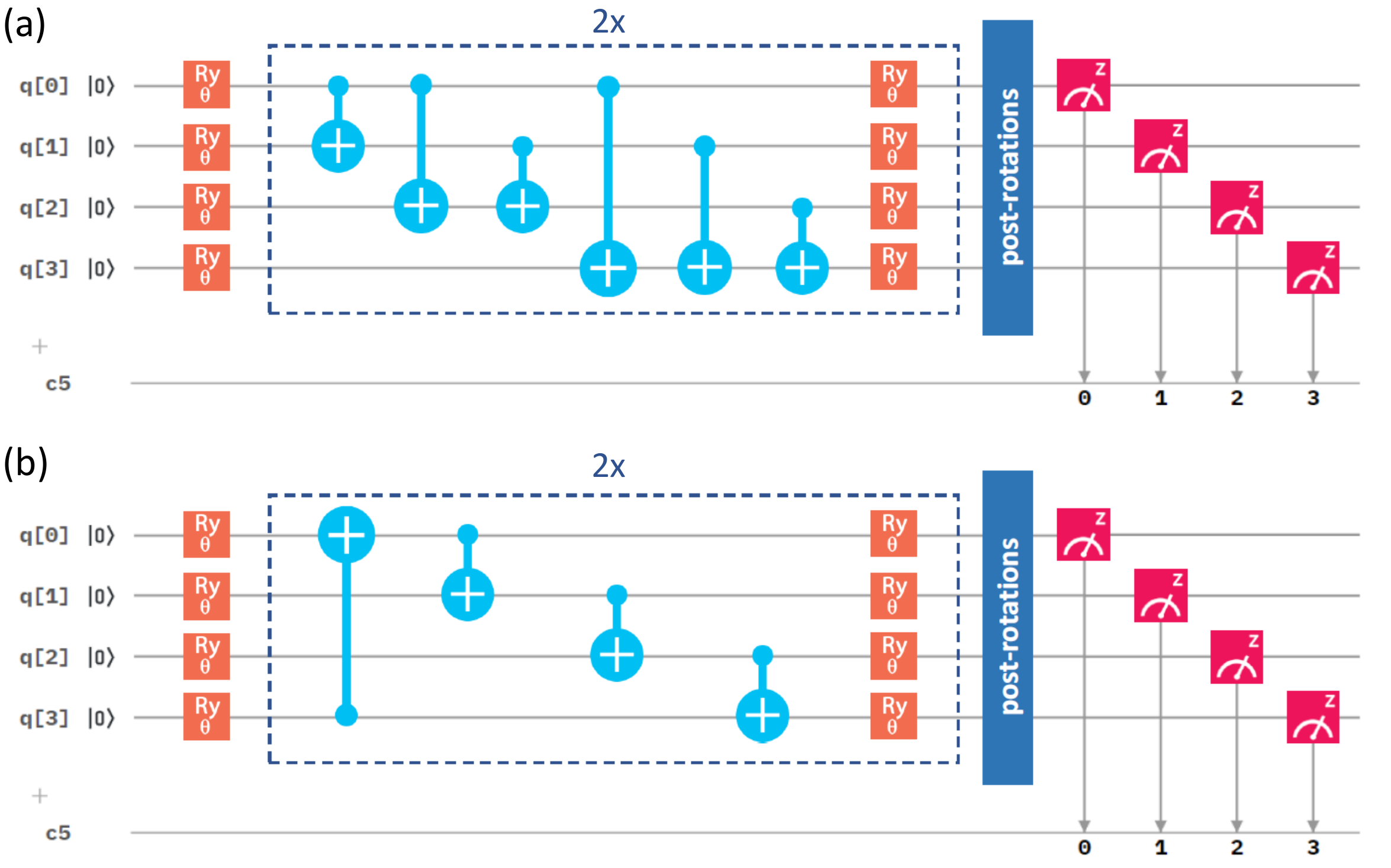}
\caption{Two additional types of qubit {entangling layers} that we have  compared with the linear one. In particular, we show a full {entangling layer} (\textbf{a}) and the circular one (\textbf{b}).}
	\label{fig-entangling-layers}
\end{figure}

Next we focus on another aspect of the circuit architecture, the {entangling layers} of qubits.
After performing all above calculations with the linear {entangling layers} (see the light-blue cNOTs in the central part of both Figs.~\ref{fig-form-RyRz} and ~\ref{fig-form-Ry}) we take into account also the circular and full {entangling layers} visualized in Fig.~\ref{fig-entangling-layers}.

The SPSA method proves to be rather insensitive to changes in the {entangling layer}, see Figs.~\ref{fig-entangling-layers}(a,b). In contrast to this robustness, the results obtained when using the COBYLA method clearly show how sensitive this method is to characteristics of the {entangling layer}.

\begin{figure}[H]
	\centering
\includegraphics[width=0.9\linewidth]{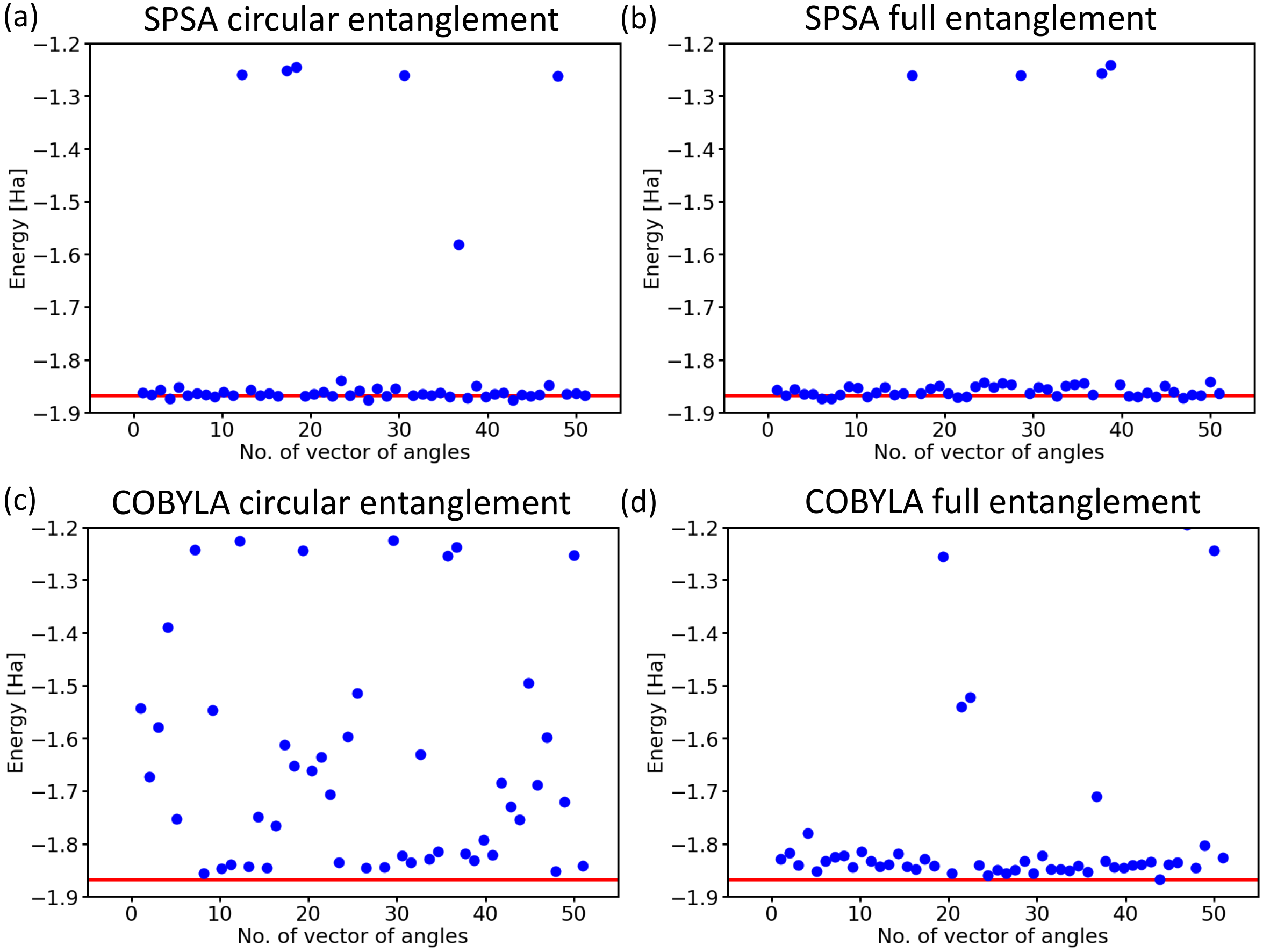}
\caption{Computed ground-state energies of the H$_2$ molecule as obtained for
50 different initial sets of random angles in the case of the SPSA optimization method with the  circular {entangling layer} (\textbf{a}) and the full one (\textbf{b}), as well as the COBYLA optimization method with either the circular {entangling layer} (\textbf{c}), or the full one (\textbf{d}).
Results are for the $R_{\mathrm{y}}$ variational form and 4096 shots.
}
	\label{fig-44}
\end{figure}


While the circular {entangling layer} significantly increases the number of erroneous results, see Fig.~\ref{fig-entangling-layers}(c), the full {entangling layer} significantly reduced the number of energies in the excited-state region from six, see Fig.~\ref{fig-43}(a), to two, see Fig.~\ref{fig-entangling-layers}(d), but three values ended in the erroneous region between the ground state and the excited ones.

\subsection{Analysis of probabilities of basis states}

{Observing results corresponding to excited states rather than the ground state, we address this issue in a detail.
As discussed in the text related to Fig.~\ref{fig-42} above, there are a few eigenvalues in the energy range between -1.2 and -1.3 Ha. Assuming that the information about phases is not available and we are limited to measure only the probabilities of basis states, we below suggest that the computed energies are sorted according to a similarity of these measured probabilities.}

Motivated by the fact that the COBYLA method seems to be more prone to provide results corresponding to the excited states, but otherwise performs quite well, we will focus our analysis on this method. Moreover,
in order to maximize the accuracy we will use below the maximum value of shots, i.e. 8192, allowed by the Qiskit package.


There are $2^{l}$ basis states for an $l$-qubit Hamiltonian and, therefore, we will analyze probabilities of the following 16 basis states:
$|$0000$\rangle$, $|$0001$\rangle$, $|$0010$\rangle$, $|$0011$\rangle$, $|$0100$\rangle$, $|$0101$\rangle$, $|$0110$\rangle$, $|$0111$\rangle$, $|$1000$\rangle$, $|$1001$\rangle$, $|$1010$\rangle$, $|$1011$\rangle$, $|$1100$\rangle$, $|$1101$\rangle$, $|$1110$\rangle$, $|$1111$\rangle$.
Examples of the measured probabilities are the following sets:

\vspace{1mm}
\noindent
set A0: (22, 4,  9,  81, 126, 0, 34, 0,  1, 2,  1, 0, 29, 0, 7880, 3)/8192 for the circuit 0

\noindent
set A1: (10,   9,    7,    23,   1220, 0,    2378, 2,    1,    11,   19,   31,   2543, 0,    1935, 3)/8192 for the circuit 1

\vspace{1mm}
\noindent
that result in the energy -1.8422 Ha and the sets

\vspace{1mm}
\noindent
set B0: (21, 2,  2,  8,  183, 0,  106, 1, 2,  0,  4,  0,  12, 0,  7839, 12)/8192 for the circuit 0

\noindent
set B1: (0, 3, 16, 10, 1111, 2, 2026, 2,  3, 1, 7, 0, 3286, 8, 1714, 3)/8192 for the circuit 1

\vspace{1mm}
\noindent
that result in the energy of -1.8464 Ha. As both energies are close to the exact value of the ground state we assume that the sets of probabilities correspond to the ground state and that differences between them are due to {the probabilistic nature of quantum states} (in the case of simulating an ideal quantum processor).
When inspecting the sets A0 and B0 for circuit 0 it seems that the probability of the basis state $|$1110$\rangle$ is {very close to} one and {nearly} zero otherwise. In such a situation it is easy to tell that the following set of measured probabilities for circuit 0

\vspace{1mm}
\noindent
set C0: (3,  2,  208, 7269, 0,  13, 28, 6,  7,  19, 2,  132, 4,  262, 7,  230)/8192
for the circuit 0

\vspace{1mm}
\noindent
is likely related to a different eigenvalue. Indeed, the corresponding energy is equal to -1.2526 Ha, i.e. a higher-lying eigenvalue corresponding to an excited state.
The situation, when the probability is {very close to} one for one of the basis states and {nearly} zero for all the others, is advantageous in the case of noise and errors containing runs because it opens the way towards identifying the probabilities that are caused by errors and noise. They can then be used in a reverse-engineering manner as parameters in noise-mitigation techniques.
Ideally, we would like to have a tool which can identify the sets A0 and B0 as similar while the set C0 as very dissimilar. Before we suggest a suitable similarity measure, it is worth discussing the probabilities related to the circuit 1 as well.

As far as the sets A1 and B1 of probabilities for the circuit 1 are concerned, the situation is not as clear as in the case of sets A0 and B0 for the circuit 0. In particular, there are four basis states with significant {outcome} probabilities but the probabilities for the same basis states in sets A1 and B1 differ by more than 10\%. {A high  number of shots is then needed to determine the probabilities reliably.} Again, it would be advantageous to have a similarity measure which identifies the A1 and B1 as similar and related to the same eigenvalue.

Considering that the probabilities are sets of non-negative values, we suggest to use the following two measures.
First, the Jaccard-Tanimoto (J-T) index, also known as the Jaccard similarity coefficient that is used for gauging the similarity and diversity of sample sets~\cite{multi-core,J-T-3,J-T-4,J-T-5,J-T-6}.
It was developed by P. Jaccard~\cite{Jaccard} and independently formulated again by T. Tanimoto~\cite{Tanimoto}. The Jaccard-Tanimoto index of two sets X and Y is defined as the ratio of intersection of the two sets over their union
$
J(X,Y) = |X \cap Y| / |X \cup Y|.
$
For two vectors \{$x_i$\}, \{$y_i$\} with all components non-negative ($x_i \geq 0$, $y_i \geq 0$) and the same length ($i = 1, ..., m$) it is evaluated as
$
J(\{x_i\}, \{y_i\}) = \sum_i \mbox{min}(x_i,y_i)/ \sum_i \mbox{max}(x_i,y_i).
$

The second measure is the scalar product of the vectors representing the set of measured probabilities. It must be emphasized that the vectors of measured probabilities have the sum of components equal to one but their length, in a vector-sense, is in general not equal to one. The length of a probability vector is equal to one only when one of the basis states has the probability close to one and all others have the probability equal to zero. It is also the maximum length in the vector sense. The other extreme case, when the length of the vector of probabilities is lowest, is the situation when all $m$ basis states have the same probability equal to 1/$m$ and the length is 1/$\sqrt{m}$.
{Therefore, we have normalized the probability vectors before we evaluated their scalar product as follows. Considering that the probabilities are squares of amplitudes of wave functions, we will use square roots of measured probabilities when evaluating the similarity by the vector product. Defined in this way, the scalar product as the second similarity measure corresponds to the upper bound of the fidelity between the states.
Below we have evaluated similarities as obtained for two sets (for circuits 0 and 1) of 500 vectors of measured probabilities. }

\begin{figure}[H]
	\centering
\includegraphics[width=0.99\linewidth]{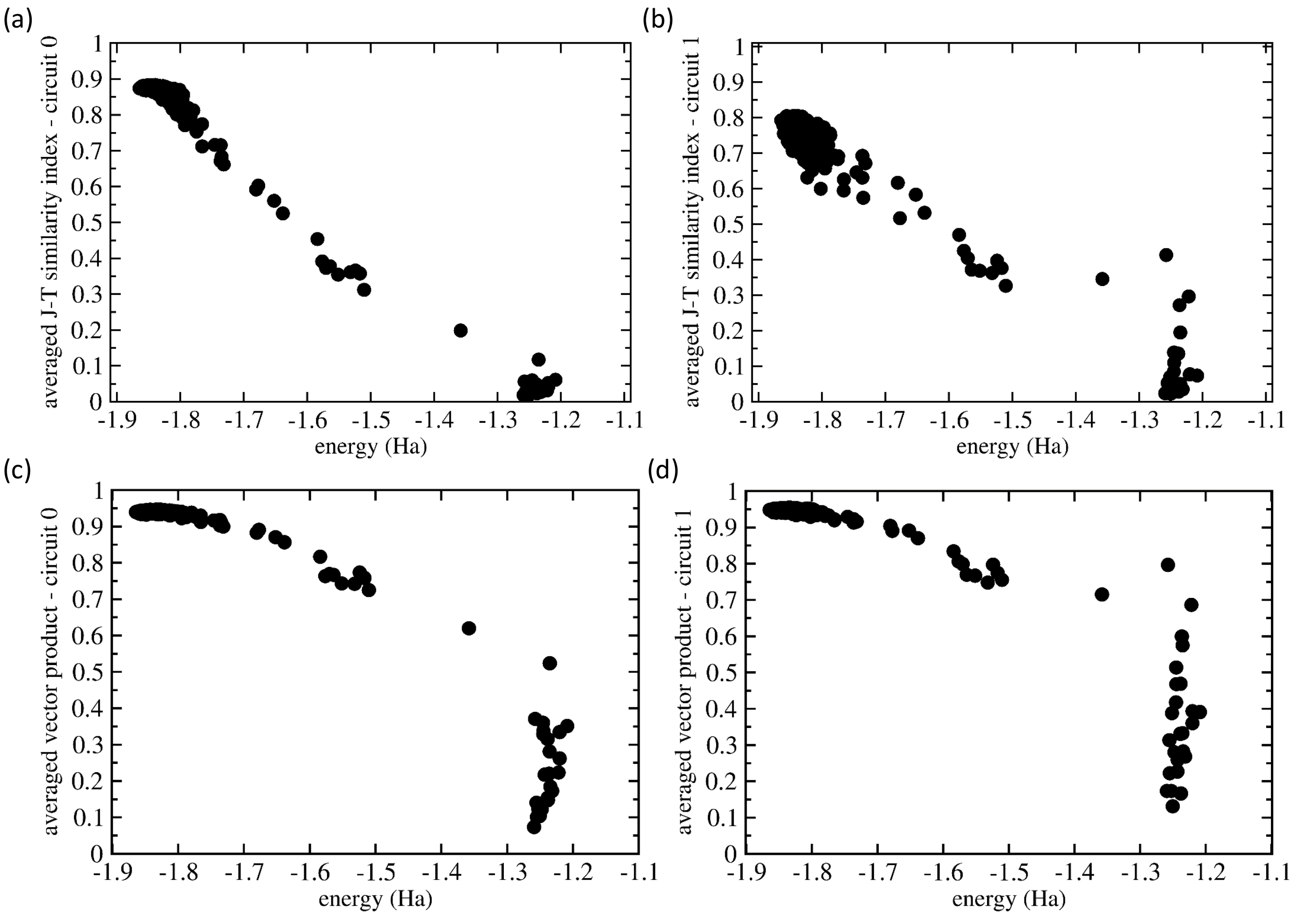}
\caption{Comparison of similarities of vectors of measured probabilities as functions of energies.
	The plotted values are averaged ones using all 500 values and evaluated using either the Jaccard-Tanimoto (J-T) similarity index (parts (a,b)) and scalar product (parts (c,d)) for both the circuit 0 (parts (a,c)) and circuit 1 (parts (b,d)). }
	\label{sim-fig}
\end{figure}

Having the two similarity measures for our analysis, we will apply them in a general manner assuming no prior knowledge of which set of probabilities are related to which eigenvalues. Our choice is motivated by the fact that future applications of quantum computers will likely be focused on systems
for which data similar to those, that we have at hand for the H$_2$ molecule, will not be available. Therefore, we determine for each vector in those two 500-member sets (for circuits 0 and 1)
J-T/scalar-product similarities of each particular vector with all 500 vectors in the set and then we
determine the average over these 500 similarity values.
Figure~\ref{sim-fig} shows the values of both the J-T similarity index (Fig.~\ref{sim-fig}(a,b)) and scalar product (Fig.~\ref{sim-fig}(c,d)) for both the circuit 0 (Fig.~\ref{sim-fig}(a,c)) and circuit 1 (Fig.~\ref{sim-fig}(b,d)).

Figure~\ref{sim-fig} shows that these averaged similarity values decrease with increasing energy
from the ground-state value. Starting with the circuit 0, as many as 450 from the 500 runs result in set of probabilities that are similar to the situation when the probability is {close to} one for the $|$1110$\rangle$ basis state and {nearly} zero otherwise. The corresponding energies cover the range from the exact value of -1.867 Ha up to -1.7 Ha. The J-T similarity index values, see Fig.~\ref{sim-fig}(a), show a weakly decreasing trend with increasing energy. In contrast, the averaged scalar products of all of these 450 probability vectors {are very close to the same value 0.95, see the results in Fig.~\ref{sim-fig}(c). As they represent an average over 450 very similar states (when the similarity values are close to 1) but about 10\% of very different states (these contribute into the average by the similarity values close to 0), the average values are not equal to 1 but only to about 0.95. }

As another extreme, about 5\% of vectors of probabilities have very low averaged similarity value (close to zero) and the corresponding energies belong into the range between -1.3 Ha and -1.2 Ha. These we identify as excited states and their vectors of probabilities are very different. Interestingly, while J-T similarity index assigns to these cases values, that are clearly close to zero without any exception, the scalar product in one case results in a value close to 0.5. The remaining 5 \% of probability vectors are characterized by energies and similarity index values that are in between the region close to the ground state on the one hand and that of excited states on the other hand. We believe that these essentially erroneous calculations should be dropped.
\newpage

The circuit 1 (see Fig.~\ref{sim-fig}(b,d)) turns out to be more complicated. The ground state possesses the probability vector in the form of a superposition of four basis states ($|$0100$\rangle$, $|$0110$\rangle$, $|$1100$\rangle$ and $|$1110$\rangle$) with rather similar probabilities (they sometimes differ by less than 1/(2$^l$) where $l$ is the number of qubits. As far as the  J-T similarity index is concerned, it performs qualitatively similarly (see Fig.~\ref{sim-fig}(b)) as in the case of circuit 0 (see Fig.~\ref{sim-fig}(a)) but some of the states, that we identified as excited, show significant non-zero values (up to 0.4). These features are even more pronounced in the case of scalar product (see Fig.~\ref{sim-fig}(d)).

\subsection{Simulations of noisy runs}

Our analysis so far has been based on simulations of ideal quantum processors that do not exhibit any noise. While these simulations currently represent a major part of published quantum-computing results,
our ultimate goal is to employ real quantum processors. The current ones based on superconductor units are, unfortunately, quite noisy, partly due to their quantum nature~\cite{Bloch-1957} but mostly due to unresolved issues related to the  technical complexity of physical realizations of quantum processors, for details see, e.g.
Ref.~\cite{ Krantz-APR-2019}. Consequently, there is a tremendous effort focused on various error-mitigation methods~\cite{Kandala2019,PhysRevX.8.031027,Cai2021quantumerror,Suchsland2021algorithmicerror,mitigation2,Geller_2021}.

\begin{figure}[H]
	\centering
\includegraphics[width=0.99\linewidth]{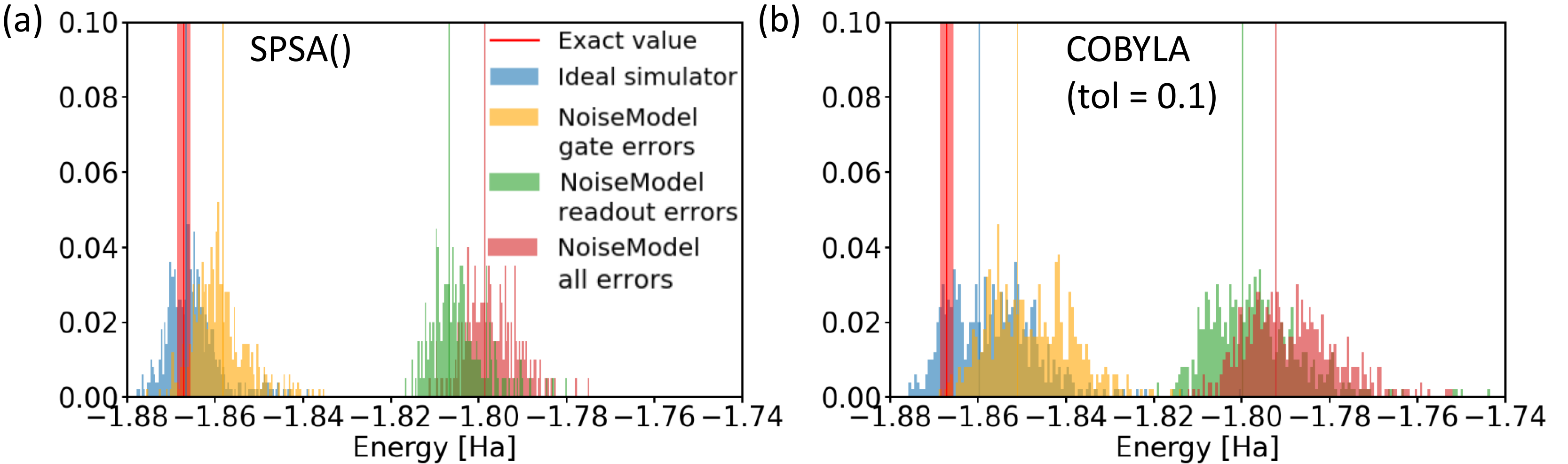}
\caption{Results of our simulations when various types of noise were included using either the SPSA (\textbf{a}), or the COBYLA (\textbf{b}) optimization method. The simulations aimed at determining
the ground-state energy of the H$_2$ molecule in the case of the 4-qubit Hamiltonian.
The red vertical thick line represents the energy region around the exact result   corresponding to the chemical accuracy. Thin vertical lines of different colors indicate the median of various sets of results. {The minimizations by the SPSA method used default parameters while those by the COBYLA method were characterized by the tolerance value equal to 0.1.}}
	\label{fig-Noise}
\end{figure}

Motivated by the above mentioned facts we extend our theoretical analysis to simulations that include noise. Its characteristic parameters are {frequently} published by the IBM for its quantum processors. Importantly, it is possible to even switch on/off different types of noise in simulations, in particular, the noise related to gate errors, readout errors and the combination of thereof. Our results are presented in Fig.~\ref{fig-Noise} for both the SPSA and COBYLA optimization methods in comparison with noise-free ideal-simulator values.

When considering all possible errors, see red-color data in Fig.~\ref{fig-Noise}, the use of both optimization methods results in a similar shift by about 0.07 Ha to higher energies. Simulations including solely gate errors or readout errors then show that the latter errors are responsible for the dominant contribution into the energy shift. In contrast, the gate errors result in a lesser part of the energy shift. Comparison of Figs.~\ref{fig-Noise} (a) and (b) indicates that the results related to the COBYLA method cover wider ranges when compared with the SPSA results but the median values are similar.

{It is worth noting that some energies in  Fig.~\ref{fig-Noise} are lower (more negative) than the exact value. This seemingly contradicts the application of the variational principle but the reason is, in fact, the impact of noise again. In particular, as the probabilities are noisy and the coefficients in the Hamiltonian (\ref{H2_4_qubits}) have both positive and negative values, a noise-related re-distribution of probabilities from positive coefficients to negative ones may results even in the energies that are lower than the exact value.}

\subsection{Experiments on real quantum processors}

\begin{figure}[H]
	\centering
\includegraphics[width=0.99\linewidth]{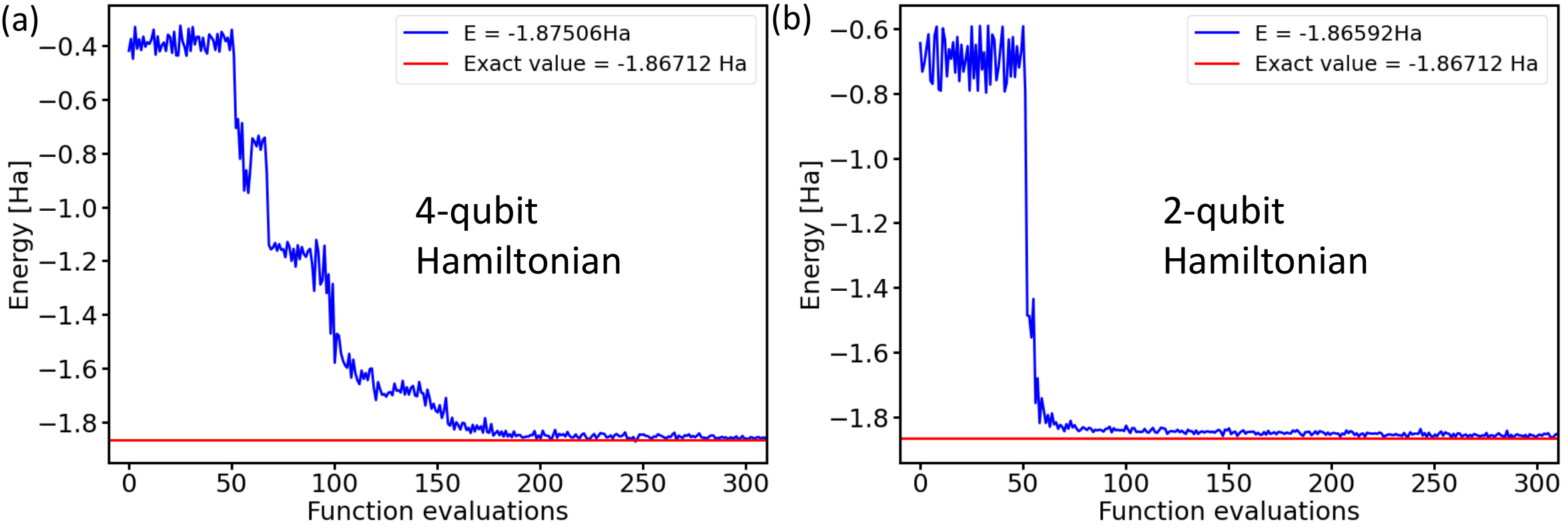}
\caption{Comparison of the calculated ground-state energies of the H$_2$ molecule in the case of 4-qubit Hamiltonian (\textbf{a}) or the 2-qubit Hamiltonian (\textbf{b}) using the $R_{\mathrm{y}}$  variational form, the linear {entangling layer}, the SPSA optimization method
and the probabilities based on 4096 shots.}
	\label{fig-47}
\end{figure}

After running noise-containing simulation, our final step consists in experiments using physical realizations of quantum processors. As these are resource-intensive, we will limit ourselves only to 2-qubit Hamiltonian. Figure~\ref{fig-47} exemplifies similarity of results for the 4-qubit and 2-qubit Hamiltonians
in the case of ideal noise-free simulations.

\begin{figure}[H]
	\centering
\includegraphics[width=0.99\linewidth]{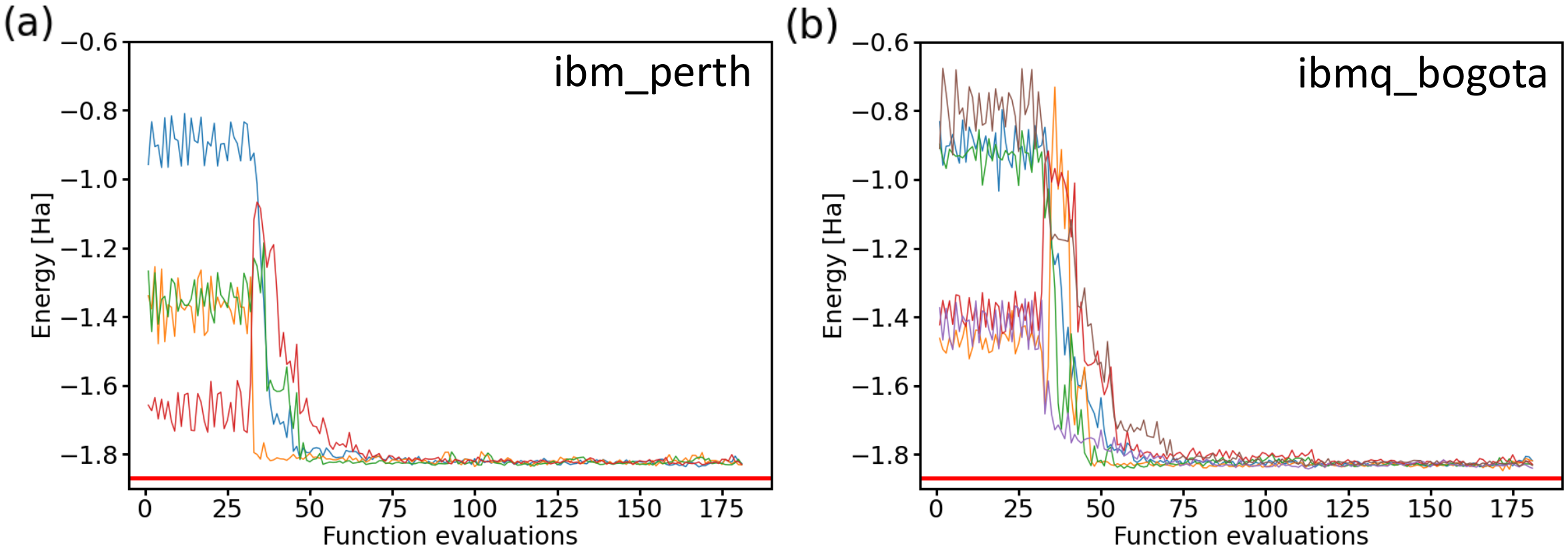}
\caption{Comparison of 10 iterative runs  (color-coded) aimed at determining the ground-state energies of the H$_2$ molecule using two IBM quantum processors. We show four runs in the case of ibm\_perth quantum processor (\textbf{a}) and six ibmq\_bogota runs (\textbf{b}) as a function of the number of function evaluations. The results were obtained for a 2-qubit
Hamiltonian, the $R_{\mathrm{y}}$  variational form, the linear {entangling layer} of qubits, the SPSA optimization method (the maxiter parameter equal to 75)
and the probabilities were based on 8192 shots.
}
	\label{fig-q-processors}
\end{figure}

After confirming similarity of results when using both 2-qubit and 4-qubit Hamiltonians, we present values obtained during 10 experiments consisting of iterative runs employing two IBM quantum processors in Fig.~\ref{fig-q-processors}. By inspecting the resulting energies it is interesting to note that while all experimental runs behave quite similarly as a function of the number of functional evaluations and they all converge to a similar energy, none of them converges to the exact value of the ground-state energy of H$_2$ molecule. Instead, there is a small shift to a slightly higher energy. The origin of this energy shift can be possibly traced to either noise and errors in real processors as indicated by our previous noise-containing simulations presented above in Fig.~\ref{fig-Noise} where a vast majority of the resulting energies was above the exact one {or, alternatively, it may be so that the correct solution was not found within a 2-qubit space of our Hamiltonian.}

\section{Discussion}


Our analysis above clearly shows that the simulations of runs of ideal quantum computers can be very sensitive to some aspects of the computational set-up. In particular, it was interesting to see that all four inspected optimization methods provide sometimes (in at least 2\% of runs) the energy of excited states instead of the ground-state energy, which we primarily search for. While these results are scattered far away from the energy of the ground state between -1.2 Ha and -1.3 Ha (see Figs.~\ref{fig-42},~\ref{fig-43}) in the case of simulations of the ideal quantum computer, the errors and noise related to real quantum processors will likely make the situation yet worse. The studied case of energies of H$_2$ molecule is very convenient because the ground-state energy is separated from the other eigenvalues by quite a broad energy range of about 0.6 Ha. Consequently, it is relatively easy to distinguish the values related to the ground state from those related to excited ones. But other systems, for which the exact ground-state energy value will not be {\it a priori} known, may have the eigenvalues closer. It can become difficult to distinguish between computed energies that (i) are noise-related to the eigenvalue that we aim at (such as the ground-state one) and (ii) those that are noise-related to other eigenvalues. {We have demonstrated how advantageous it is then to exploit measured probabilities of different basis states. In particular, we have analyzed similarities between the sets of measured probabilities by employing either the Jaccard-Tanimoto similarity index, or the scalar product (preceded by a square-root normalization of measured probabilities).}

{
We suggest that the similarity analysis becomes a topic of future studies, possibly employing other tools, such as the cluster analysis, because of its advantages}. First, a similarity analysis of measured probabilities can support the assessment of the computed energies and our results indicate that it can be able to distinguish cases that are just noisy from those that are related to other eigenvalues, such as higher-energy excited states. Second, the two similarity measures that we employed, the Jaccard-Tanimoto similarity index and the scalar product, perform clearly better when the ground-state energy is associated with one of the basis states having the probability close to one while the others have it zero. Third, various similarity measures have specific advantageous and disadvantageous characteristics. For example, when identifying states related to the ground state, the similarity analysis based on the scalar product seems to be more successful than the one employing the J-T index. On the other hand the former tends to overrate the similarity of the excited and completely erroneous states.

\section{Conclusions}

Motivated by the progress of quantum computing in quantum chemistry we have performed an extensive series of simulations of quantum-computer runs that were aimed at inspecting practical aspects of these calculations. In order to compare the performance in the case of different set-ups, the ground-state energy of hydrogen molecule has been chosen as a benchmark example of a system for which accurate solutions exist in literature. Employing variational quantum eigensolver (VQE) we have analyzed the impact of various computational technicalities including (i) the choice of optimization methods (COBYLA, SPSA, Nelder-Mead and Powell's), (ii) the architecture of quantum circuits (linear, circular and full  {type of qubit entangling layers} and $R_{\mathrm{y}}R_{\mathrm{z}}$  and $R_{\mathrm{y}}$ variational forms), as well as (iii) different types of noise (gate errors and readout errors) when simulating real quantum processors. We have also performed a series of experimental runs as a complement to our simulations.

The SPSA and COBYLA methods have significantly outperformed the Nelder-Mead and Powell optimization methods. The results obtained when using the $R_{\mathrm{y}}$ variational form have been found better than those when the $R_{\mathrm{y}}R_{\mathrm{z}}$ form was used. In particular, a statistical distribution of computed energies is spanning a narrower range and closer to the exact value. Further, the iterative runs are less likely to lead to excited states instead of the ground state that we primarily aim at. The choice of an optimum {entangling layer} is interlinked with the choice of the optimization method. For example, the performance of the COBYLA method has been found to worsen when the circular {entangling layer} is used while the full {entangling layer} improves its performance. In contrast to this sensitivity of the COBYLA method, the SPSA method turns out to be quite robust with respect to the {entangling layer(s)} type.

Further, all four inspected optimization methods sometimes lead to an energy that corresponds to an excited state rather than the ground state. We show that a similarity analysis of measured probabilities using the Jaccard-Tanimoto similarity index or the scalar product can provide a very useful insight into these cases.
The similarity analysis of measured probabilities can support the assessment of the computed energies and our results indicate that it can be able to tell cases that are just noisy from those that are related to other eigenvalues, such as excited states with higher energies.

Importantly, both similarity measures perform clearly better when the ground-state energy is associated with the situation when one of the basis states has the probability equal to one while the others have it zero. In contrast to the above mentioned characteristics that are common to both used similarity measures, there are significant differences between them, too. In particular, when identifying states related to the ground state, the similarity analysis based on the scalar product seems to be more successful than that employing the J-T index. On the other hand the former tends to overrate the similarity of the excited and completely erroneous states.

\authorcontributions{Writing—Original Draft Preparation, visualization, I. M., M. F., M. Pi., M. Pl., M. S. and M. {\v S}.; Conceptualization, Methodology, M. Pi., M. Pl., M. S. and M. {\v S}.; Writing—Review \& Editing, I. M., M. F., M. Pi., M. Pl., M. S. and M. {\v S}.; Resources, Project Administration, Funding Acquisition,
M. F. and M.{\v S}.; Supervision, M.{\v S}. and M. Pl.}

\funding{We gratefully acknowledge financial support from the Grant Agency of Masaryk University in Brno, Czech Republic, within an interdisciplinary research project No. MUNI/G/1596/2019 entitled "Development of algorithms for application of quantum computers in electronic-structure calculations in solid-state physics and chemistry". M.Pi. and M. Pl also acknowledge funding from VEGA Project No. 2/0136/19.}

\institutionalreview{{Not applicable.}}

\informedconsent{{Not applicable.}}

\dataavailability{{The data presented in this study are available on request from the corresponding author.}}

\acknowledgments{Computational resources were provided by the Ministry of Education, Youth and Sports of the Czech Republic under the Projects e-INFRA CZ (ID:90140) at the IT4Innovations National Supercomputing Center and e-Infrastruktura CZ (e-INFRA LM2018140) at the MetaCentrum as well as the CERIT-Scientific Cloud (Project No. LM2015085), all granted within the program Projects of Large Research, Development and Innovations Infrastructures. M.F., I.M. and M.{\v S}. acknowledge the support provided by the Czech Academy of Sciences (project No. UFM-A-RVO:68081723). We acknowledge the use of IBM Quantum services for this work. The views expressed are those of the authors, and do not reflect the official policy or position of IBM or the IBM Quantum team. We acknowledge the access to advanced services provided by the IBM Quantum Researchers Program.}

\conflictsofinterest{The authors declare no conflict of interest.}







\bibliography{sample}

\end{document}